# Active control of surface plasmon resonance in MoS$_2$-Ag hybrid nanostructures

Shuai Zu [1†], Bowen Li[1†], Yongji Gong[2], Pulickel M. Ajayan[2], and Zheyu Fang[1,*]

[1]School of Physics, State Key Lab for Mesoscopic Physics, Peking University, Beijing 100871, China.

[2]Department of Materials Science and NanoEngineering, Rice University, 6100 Main Street, Houston, Texas 77005, United States.

*Email: zhyfang@pku.edu.cn

†These authors contributed equally to this work.

**Molybdenum disulfide (MoS$_2$) monolayers have attracted much attention for their novel optical properties and efficient light-matter interactions. When excited by incident laser, the optical response of MoS$_2$ monolayers was effectively modified by elementary photo-excited excitons owing to their large exciton binding energy, which can be facilitated for the optical-controllable exciton-plasmon interactions. Inspired by this concept, we experimentally investigated active light control of surface plasmon resonance (SPR) in MoS$_2$-Ag hybrid nanostructures. The white light spectra of SPR were gradually red-shifted by increasing laser power, which was distinctly different from the one of bare Ag nanostructure. This spectroscopic tunability can be further controlled by near-field coupling strength and polarization state of light, and selectively** **applied to the control of plasmonic dark mode. An analytical Lorentz model for photo-excited excitons induced modulation of MoS$_2$ dielectric function was developed to explain the insight physics of this SPR tunability. Our study opens new possibilities to the development of all-optical controlled nanophotonic devices based on 2D materials.**

Molybdenum disulfide (MoS$_2$) monolayers have attracted tremendous interests in recent years because of their novel direct bandgap electronic structure[1,2], versatile optical transitions[3,4] and tunable mechanical properties[5,6]. Tightly bound excitons in the MoS$_2$ monolayers make it good candidate for both fundamental physical studies and optoelectronic applications, and provides significant opportunity to realize light-emitting devices that operated at the visible spectral range[7-11]. The inherent atomic thickness of MoS$_2$ monolayers restricts the light-matter interaction, which results in poor light emission and limited its absorption behavior. A wide variety of platforms including photonic crystals[9], high Q micro-cavities[12,13], flat metal surfaces[14] and plasmonic nanoantennas[15-17], have been developed to boost the coupling between electromagnetic fields and excitons. Excitonic photoluminescence (PL) properties of MoS$_2$ monolayers were determined by its band structure and carrier density[1,18]. Great efforts have been devoted to extrinsically control the properties of MoS$_2$ excitons, including gate-bias charging[18,19], physical adsorption[20], chemical doping[21,22] and plasmonic hot electron doping[23,24]. However, to date, few studies have used laser to modify the excitonic properties of $MoS_2$ monolayers and further control the coupling between electromagnetic fields and excitons.

Collective electron oscillations in metallic nanostructures, known as surface plasmon resonance (SPR), which exhibits versatile talents for nanoscale light confinement and strong localized electromagnetic field enhancement. The abundant electromagnetic modes induced by SPR, show great potential applications in nanophotonics, such as optical force[25], light harvesting[26,27] and bio-sensing[28]. Importantly, the electromagnetic fields associated with the localized surface plasmon (LSP) modes of the metallic nanoparticles are confined to deep subwavelength volumes in all three dimensions, which provides the way to light-focusing in nanoscale and coupling enhancement with excitons. The optical properties of the nanostructure, e.g. the resonance frequency and polarization, can be easily tailored by changing its geometry, constituent material and dielectric environment[29,30]. This enables active control of optical properties of bulk materials and easily integration with the exciton system to attain strong exciton-plasmon coupling. The optical properties of both 2D materials[31] and metallic nanostructures[9,32-35] are particularly sensitive to the electromagnetic environment, which provides a unique opportunity to actively control their interactions.

In this work, we demonstrate that the optical response of $MoS_2$ monolayers can be effectively modified by the photo-excited excitons[36] which can be used to active control of surface plasmon resonance (SPR) in $MoS_2$-Ag hybrid nanostructures. We show that when the hybrid structure was illuminated by the incident laser, the plasmon resonance could be monotonically red-shifted depending on the incident power. This spectroscopic tunability that related with the near-field coupling strength and polarization state of white light source could be effectively controlled by changing the Ag nanostructure geometry. These results were further confirmed and explained by the plasmon-exciton coupling model associated with the photo-excited excitons induced modulation of $MoS_2$ dielectric function.

**Results**

**Design of $MoS_2$-Ag hybrid nanostructures and basic concept of laser controlled SPR.** The $MoS_2$-Ag hybrid nanostructures were fabricated by patterning designed Ag-disk structures onto the surface of $MoS_2$ monolayers, as shown in Fig. 1a. In order to increase the light absorption, a Ag-mirror substrate was prepared by depositing Ag (100nm) and $SiO_2$ (50nm) on the $SiO_2$/Si substrate by thermal evaporation. The chemical vapor deposition (CVD) grown $MoS_2$ monolayers were transferred onto the substrate by using the polymethyl methacrylate (PMMA) nanotransfer method. Then Ag nanostructures were fabricated on $MoS_2$ via E-beam lithography and electron beam evaporation. The thickness of all the structures is 30nm. Ag was chosen because of its strong plasmonic resonance and the relatively low dissipation in the visible range. The reflection and absorption spectra were collected by using our home-built microscope with white light illumination and laser excitation (See Methods and Supplementary Fig. 1). When excited by 532 nm laser, $MoS_2$-Ag hybrid nanostructures show active control of plasmonic resonance. To intuitively understand this concept, we can consider localized surface plasmon (LSP) of Ag and the exciton of $MoS_2$ as two coupled classical oscillators, as shown in Fig. 1b. Photo-excited excitons can effectively modify the oscillation strength of $MoS_2$ exciton. Substantially, the oscillation of LSP was changed through the exciton-plasmon coupling, which can be verified in the LSP absorption spectra as shown in Fig. 1c. The excitons generated by the laser excitation change the dielectric environment of the hybrid structure, which leads to the LSP resonance shifting when the incident laser is shining on.

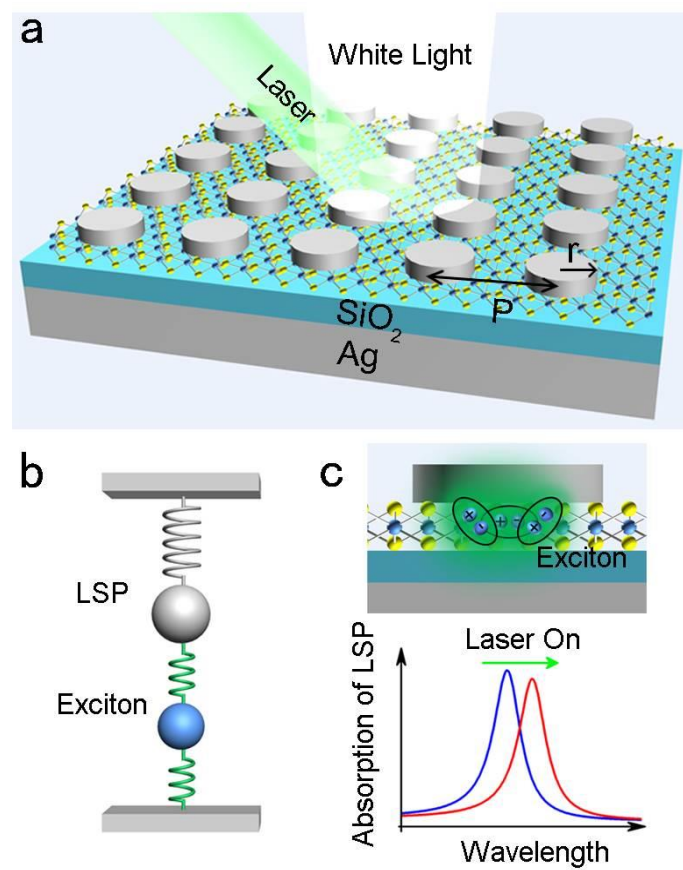

**Figure 1 | Schematic of the $MoS_2$-Ag hybrid nanostructure and coupled oscillator model for intuitive explanation.** (**a**) Periodic $MoS_2$-Ag hybrid nanostructures under 532 nm laser excitation, the absorption spectra of SPR were measured independently using a white light source. The substrate consists of a 100 nm Ag and 50 nm $SiO_2$ layer. Ag disks array with radius $r$ = 60 to 80nm and period $p$ = 445nm were fabricated on the $MoS_2$ monolayers. (**b**) Two-oscillator model for the LSP (gray ball) and exciton (blue ball) coupling. The oscillation strength of $MoS_2$ exciton can be controlled by the laser excitation. (**c**) LSP absorption spectra show a clear red-shifting with the change of the $MoS_2$ dielectric permittivity that modulated

by the photo-excited excitons.

**Characterization of MoS$_2$-Ag hybrid nanostructures.** The optical image of MoS$_2$ monolayers on the SiO$_2$/Ag substrate was shown in Fig. 2a. The MoS$_2$ layer shows orange because of the color contrast on the designed substrate. The edge length of the triangle MoS$_2$ layer is about 25 μm. Figure 2b is the SEM image of Ag disk array with radius of 60 nm and period of 445 nm that patterned at the edge area of the MoS$_2$ layer. Clear color contrast makes it easy to identify the MoS$_2$ region. Figure 2c shows the difference in spectra of MoS$_2$ monolayers and MoS$_2$ integrated with Ag disk array. The PL spectrum of pristine MoS$_2$ (red curve) shows strong PL emission at ~680 nm, while the PL of MoS$_2$-Ag hybrids (blue curve) exhibits nearly twice of magnitude enhancement than that of MoS$_2$ as shown in Fig. 2c. The inset of Fig. 3c shows the Raman spectra of pristine MoS$_2$ and hybrid MoS$_2$-Ag film. For the pristine MoS$_2$, the in-plane ($E_{2g}^1$) and out-of-plane ($A_{1g}$) modes appear at 389 cm$^{-1}$ and 412 cm$^{-1}$, respectively. In the hybrid Ag-MoS$_2$ structure, two Raman modes shift to 390 cm$^{-1}$ and 412 cm$^{-1}$, and show intensity enhancement with the near-field interactions between LSP and MoS$_2$. The enhanced emission at room temperature is due to the local field concentration at the position of the emitter arising from the LSP resonances of the disk array, suggesting weak coupling between the MoS$_2$ excitons and LSPs. The absorption spectrum was also measured, which displays characteristic peaks of A- and B-excitons at ~660 nm and ~610 nm[8,15,37] (See Supplementary Fig. 2).

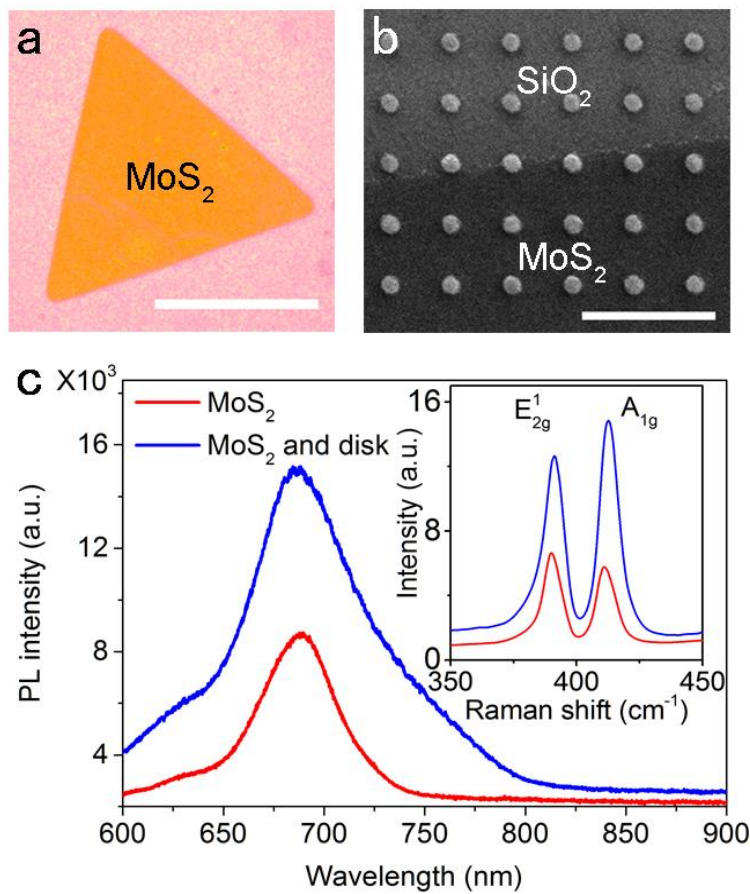

**Figure 2 | Characterization of MoS$_2$ monolayers and MoS$_2$-Ag disk hybrid nanostructures.** (**a**) Optical image of MoS$_2$ monolayers under a 100× objective lens (scale bar is 15 μm). (**b**) The SEM image of MoS$_2$-Ag disk hybrid nanostructures with disk radius of 60 nm and period of 445 nm (scale bar is 1μm). (**c**) The PL spectra of MoS$_2$ monolayers (red curve) and hybrid nanostructures (blue curve) on the SiO$_2$/Ag substrate. The MoS$_2$ monolayers display strong light emission peak at ~680 nm. Inset: corresponding Raman spectra of the two samples.

**Laser control of surface plasmon resonance.** A series of Ag disk arrays were fabricated to systematically investigate the exciton-plasmon interactions in MoS$_2$-Ag hybrid nanostructures. Because of the size dependent resonant coupling between the light and LSPs of Ag disks, different SPRs from ~700 to ~900 nm were observed (See Supplementary Figs 3-5), making it convenient to discuss wavelength-dependent coupling between MoS$_2$ excitons and LSPs. Figure 3a shows the variation of absorption peaks of MoS$_2$-Ag nanostructure (disk radius of 60 nm) with 532 nm laser excitation. With the increasing laser power from 0 to 4 mW, the absorption intensity decreases and peak position red-shifts from 695 to 720 nm. Additional experiments by using Ag disks (radius of 60 nm) directly fabricated on the same $SiO_2$/Ag substrate were performed. As shown in Fig. 3b, the absorption peak located at 681 nm, and keeps unchanged with the increasing of the laser power, which confirms that our measured spectroscopic tunability is the result of the $MoS_2$-Ag hybridization rather than the bare Ag plasmon coupling. The oscillation strength of $MoS_2$ excitons can be modulated by the primary photo-excited excitons generation. The changing of the absorption spectra of $MoS_2$-Ag structure with increasing power reveals the modulation of dielectric permittivity of $MoS_2$ and the coupling between excitons and LSPs.

To further investigate wavelength-dependency of the spectroscopic tunability, the absorption spectra of different $MoS_2$-Ag nanostructures were measured as shown in Fig. 3c and 3d, respectively. For the 65 nm-sized disk, the absorption intensity decreases weakly with the increasing of the power, and the peak position red-shifts from 711 to 724 nm which is less than the shifting value of the 60 nm-sized disk array. When the disk radius increased to 70 nm, the changing of peak position is even lower (~7 nm), and the absorption intensity is nearly unchanged with the power increasing. From Fig. 3, we can see that the spectroscopic tunability decreases when the LSP resonance is away from the $MoS_2$ absorption peak (~660 nm). For even larger disks (radius of 75 and 80nm), the shift of LSP is further weakened until the incident laser loses its spectroscopic tunability as shown in Supplementary Fig. 6. These results show that the exciton-plasmon coupling strongly depends on the spectral overlap of MoS$_2$ excitons and the LSP modes, which agree well with our finite element method (FEM) simulations (See Supplementary Note 1 and Supplementary Fig. 7). When the LSP resonant energy is tuned close to the exciton energy, the near-field coupling strength of MoS$_2$ and Ag disk is dramatically increased, and enables more sensitive spectroscopy under laser control. As a consequence, control of the near-field coupling strength is the key to modulate the exciton-plasmon hybridization.

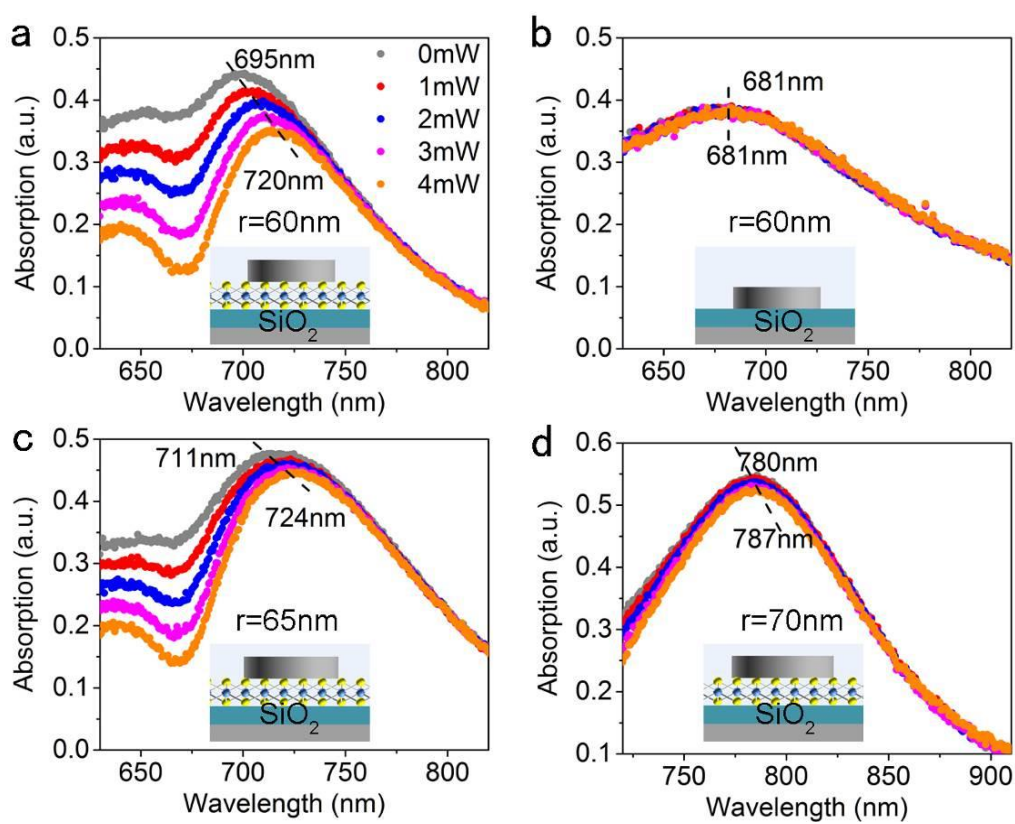

**Figure 3 | Absorption spectra of MoS$_2$-Ag hybrids with different sized Ag disk under various laser powers.** (**a**) Absorption spectra of MoS$_2$-Ag hybrids with disk radius of 60 nm, where the LSP resonance red-shifts with the increasing of the laser power. A total shifting of ~25 nm was obtained as the black dotted line shown. (**b**) Absorption spectra of Ag disk array with radius of 60 nm directly fabricated on the bare SiO$_2$/Ag substrate. The LSP resonance shows no moving tendency with the laser power increasing. (**c,d**) Absorption spectra of MoS$_2$-Ag hybrids with Ag disk radius of (**c**) 65 nm and (**d**) 70 nm. The tunability is weakened because of the non-resonant coupling. The legend for different incident laser powers is shown in (**a**).

Incident polarization state can also be used to control the near-field coupling strength of the excitons and LSPs, and switch on/off the spectroscopic tunability. As shown in Fig. 4a, Ag nanorod array was designed and fabricated on the $MoS_2$ monolayers to investigate the polarization-dependent coupling effect. The length and width of nanorod is 105 and 60 nm, and the period of array is 445 nm. The electric near-field distributions of the Ag nanorod array were simulated using FEM at resonance wavelength of 740 nm (See Supplementary Fig. 8) that corresponds to the longitudinal LSP mode with the incident polarization along the long (Fig. 4b) and short (Fig. 4c) axis of the rod. From the simulation, we can see that the electric field enhancement of horizontally polarized excitation is about 4 times larger than the one of vertical excitation, making it possible to control the spectroscopic tunability by using the light polarization state. As a result, under the horizontally polarized excitation, the absorption intensity of the $MoS_2$-Ag hybrids decreases dramatically with the laser power increased from 0 to 5 mW, and the peak position red-shifts from 749 to 770 nm (Fig. 4d). When the polarization changes to vertical (Fig. 4e), the longitudinal plasmon resonance of the Ag nanorod is switched off, and the absorption spectra only present weak coupling and spectroscopic tunability.

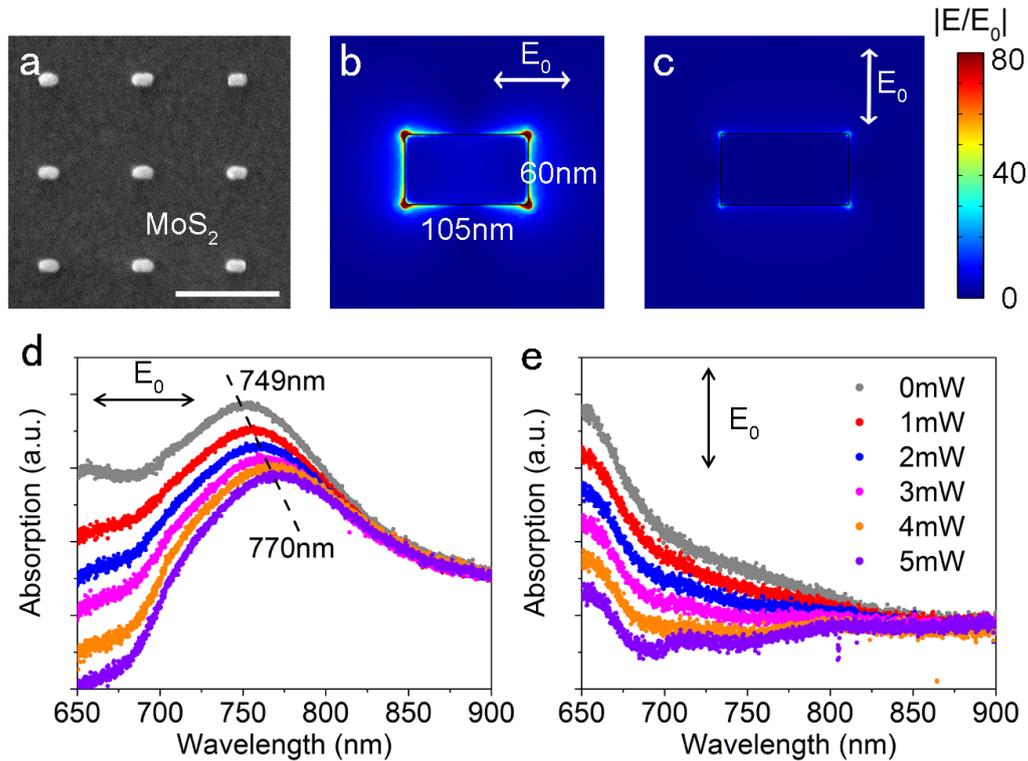

**Figure 4 | MoS$_2$-Ag nanorods hybrid nanostructure.** (**a**) The SEM image of the Ag nanorod array on the MoS$_2$ monolayer with the length of 105nm, the width of 60nm and period of 445nm. The scale bar is 500nm. (**b,c**) The simulated electric field distribution of Ag nanorod array in horizontally (**b**) and vertically (**c**) polarized lights. The white arrow shows the polarization direction. (**d,e**) The absorption spectrum of MoS$_2$-Ag nanorod hybrid film under the same incident polarization as (**b**) and (**d**). The legend of different laser powers is shown in (**e**).

This exciton-plasmon coupling can also be used to realize selectively control of plasmonic dark modes. As shown in Fig. 5a, Ag quadrumer structure consists of four nanorods was fabricated on the MoS$_2$ monolayers with SiO$_2$/Si substrate, which provides strong near-filed enhancement and high sensitivity to the change of local refractive index[38]. The length of the top and bottom nanorods is 160 nm, while for the middle two is 85 nm. The width of all the rods keeps as 60 nm, and the gap between neighboring ones is 30 nm. Plasmonic Fano resonance can be generated at wavelength of 731 nm when the polarization of incident light is along the long axis of the rod, as shown in Supplementary Fig. 9. Surface charge distribution of the quadrumer structure is shown in Fig. 5a, where the opposite alignment of electric dipoles of the middle and outer nanorods confirms this plasmonic Fano interference. Electric field distribution of this $MoS_2$-Ag quadrumer at the resonance was also plotted as Fig. 5b with the field enhancement confined at the gap and cutting-edge of the nanorods. When the incident laser power is increased from 0 to 6mW, the resonance dip red-shifts from 708 to 744 nm (Fig. 5c), showing the largest spectroscopic tunability which attributed to the strong near-filed coupling of the plasmonic dark mode and the $MoS_2$ exciton, and the high localized dielectric sensitivity of the quadrumer structure. As a result, plasmonic dark modes, such as Fano resonance, can be more efficiently and selectively controlled in comparison with the bright dipolar modes that supported by the Ag disk and nanorod array.

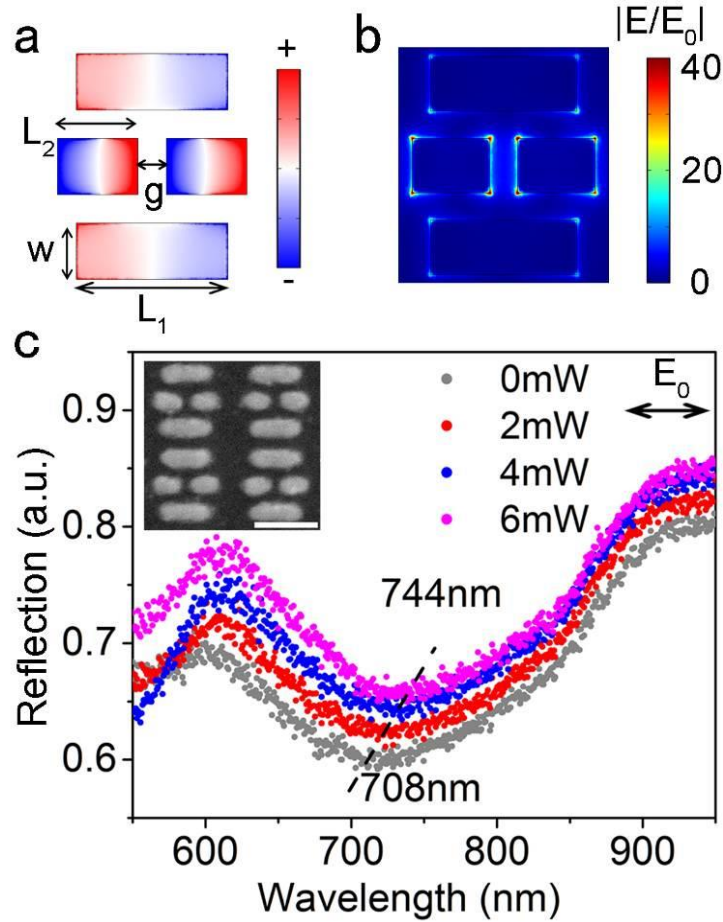

**Figure 5 | The MoS$_2$-Ag quadrumer nanostructure.** The FEM simulated surface charge (**a**) and electric field (**b**) distribution at the Fano resonance. The top and bottom nanorod: the length $L_1$ = 160nm, the width $W$ = 60nm. The middle nanorod: the length $L_2$ = 85nm, the width $W$ = 60nm. All the gap is $g$ = 30nm. Period is 280nm. (**c**) The reflection spectra of MoS$_2$-Ag quadrumer nanostructure in the horizontally polarized white light illumination under different incident laser powers. Inset: The SEM image of the Ag quadrumer array on monolayered MoS$_2$. The scale bar is 200nm.

**Analytical calculations of photo-induced SPR shifting.** To further analyze the observed optical tunability of MoS$_2$-Ag hybrid nanostructures under laser excitation, FEM simulations were performed to calculate the absorption and reflectionspectra. Both photo-excited excitons and free carriers doping have been proved can contribute to the dielectric permittivity change of the MoS$_2$ monolayers[24,36,39]. Despite the blue-shifting that induced by the surface charge doping, the obvious spectroscopic red-shifting that we recorded in the experiment demonstrates that the photo-excited excitons are the primary contribution for the measured absorption spectrum, which can be modeled by using a simple Lorentz oscillator description[36,39-41]. The dielectric constant $\tilde{\varepsilon}_{MoS_2}(\omega)$ of MoS$_2$ can be expressed as

$$\tilde{\varepsilon}_{MoS_2}(\omega) = \tilde{\varepsilon}_{\exp}(\omega) + \frac{f}{\omega_0^2 - \omega^2 - i\omega\gamma} \quad (1)$$

where $\tilde{\varepsilon}_{\exp}(\omega)$ is the experimental MoS$_2$ dielectric response[42], $f$ is the equivalent oscillator strength, $\gamma$ (118 meV) and $\omega_0$ (1.884 eV) are the bandwidth and transition energy of the MoS$_2$ exciton. Oscillator strength $f$ is direct proportional to the density of excitons which depends on the laser power $P$ and corresponding absorption cross section $A$, as $f \propto A \times P$. When laser power and absorption cross section increase, the dielectric constant of MoS$_2$ increases with the $f$, resulting in red-shifting of LSP resonance which was predicted by our coupled oscillator model in Fig. 1b.

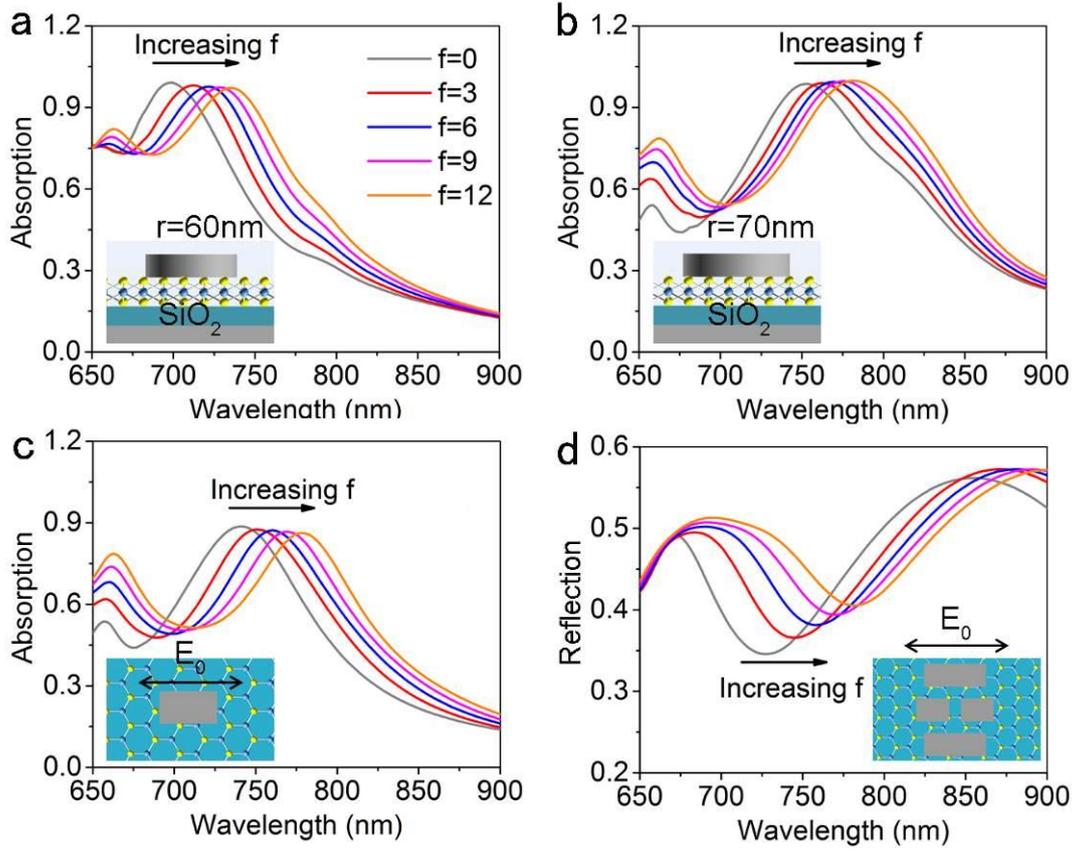

**Figure 6 | Simulated absorption and reflectionspectra of disks, nanorod and quadrumer MoS$_2$-Ag hybrid nanostructures by changing oscillator strength *f*.** (**a**) *r* = 60nm (**b**) *r* = 70nm (**c**) nanorod with the length of 105nm and the width of 60nm under horizontally polarized light illumination and (**d**) quadrumer under horizontally polarized light illumination. The dielectric constant of MoS$_2$ increases as *f* increasing causing the red-shifting of the LSP resonance. The legend of different oscillator strength *f* is shown in (**a**).

Figure 6 shows the absorption and reflectionspectra of disks, nanorod and quadrumer MoS$_2$-Ag hybrid nanostructures by changing *f*. As expected, all the LSP resonances red-shift as *f* increasing, which is consistent well with the experiment results. For the 60 nm-sized disk array (Fig. 6a), the LSP resonance red-shifts from 699 to 735 nm with the *f* increasing from 0 to 12. When the disk size increases to 70 nm (Fig. 6b), because the near-field coupling strength is weakened, the spectroscopic red-shifting decreases. Larger sized disks array with different radius were also calculated and show similar decreasing tendency (See Supplementary Fig. 10). For the nanorod and quadrumer structures, the calculated peak positions for both plasmonic bright and dark modes present exactly the same red-shifting phenomena, which further confirms our theoretical oscillator coupling prediction.

**Discussion**

In summary, we have shown that laser-induced dielectric constant change of $MoS_2$ can modify its optical properties and control the exciton-plasmon coupling in $MoS_2$-Ag hybrid nanostructures. The localized surface plasmon of Ag nanostructures is found to exhibit both resonance red-shifting and intensity change that depend on the laser power and light polarization state, and inherently controlled by the near-field plasmon-exciton coupling strength. This spectroscopic tunability can be selectively applied to the control of plasmonic dark modes. The largest spectral shifting (~36 nm) was observed with the plasmonic Fano resonance close to the $MoS_2$ exciton peak. The capable of active control of exciton-plasmon coupling would open new opportunities in novel nanophotonic devices based on 2D materials, such as plasmonic circuits and optical modulators at the nanoscale.

**Methods**

**Optical Measurement**

The Raman, reflection and PL spectra were measured by a home-built optical microscope equipped with a 100× objective lens (Fig. S1). The white light source (Dolan-Jenner MI-150 with fiber illumination) was linearly polarized with a polarizer and focused onto the back focal plane of the objective for the reflection spectrum measurement. The incident laser was combined with another optical path and focused onto the MoS$_2$-Ag hybrid structures for the PL excitation. The surface plasmon resoance and reflection spectra of the hybird structures were collected by the spectrometer (iHR550, Horiba Co.) and corrected by using a standard white calibration (Edmund).

**Numerical Simulations**

Full-field electromagnetic wave simulations were performed using the finite element solver (COMSOL Multiphysics). A unit cell of the investigated structure was simulated using periodic boundary conditions along the x- and y-axes and perfectly matched layers along the propagation of electromagnetic waves (z-axis). Plane waves were launched and illuminated the unit cell along the -z direction. In the simulations, we used Palik data[43] for the Ag and Si complex refractive indices and the refractive index of SiO$_2$ is 1.45. The dielectric permittivity of MoS$_2$ monolayers was extracted by parametrizing experimental data into a band and exciton transition (Brunaur-Emmett-Teller) model as reported in previous works[24,42].

**Acknowledgements**
This work was supported by the National Basic Research Program of China (973 Program, Grant No. 2015CB932403), National Science Foundation of China (Grant No. 11374023, 61422501 and 61521004), and Beijing Natural Science Foundation (Grant No. L140007). Foundation for the Author of National Excellent Doctoral

Dissertation of PR China (Grant No.201420), National Program for Support of Top-notch Young Professionals.

**Author contributions**

**Competing financial interests:** The authors declare no competing financial interests.

# Supplementary Figures

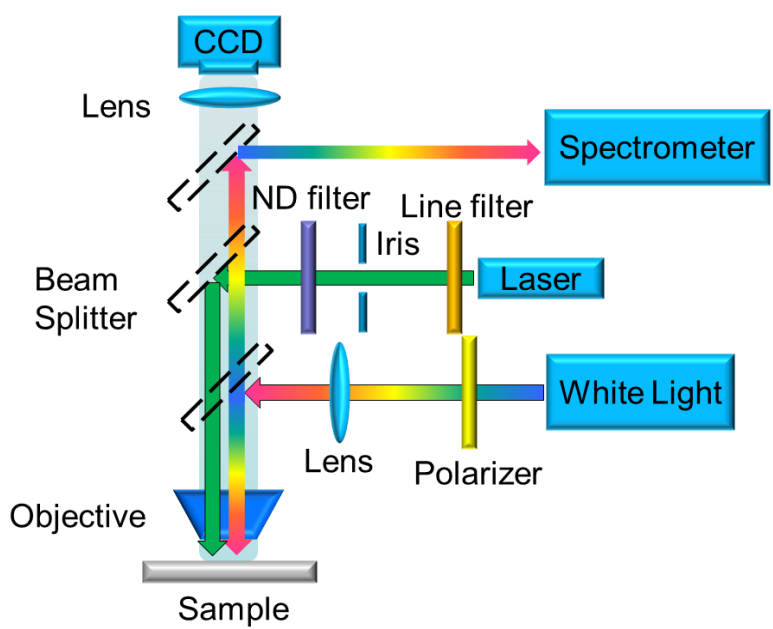

**Supplementary Figure 1.** Schematic of the optical setup for the absorption, reflection, photoluminescence and Raman spectra measurement. ND filter: Neutral density filter.

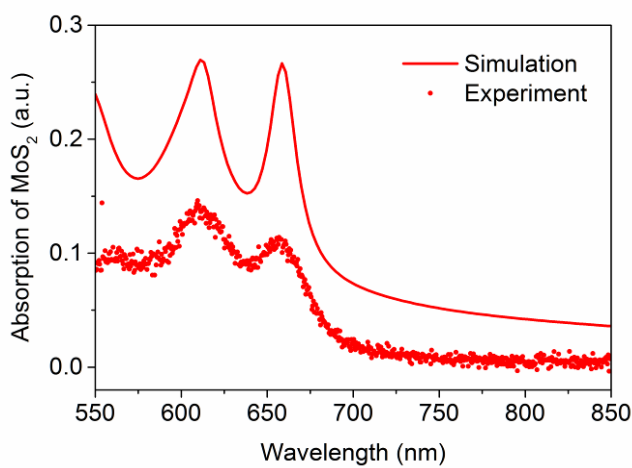

**Supplementary Figure 2.** The experimental (red dot) and simulated (red line) absorption spectrum of $MoS_2$ monolayers on $SiO_2$/Ag substrate. For pristine $MoS_2$, the absorption peak of A- and B-excitons located at ∼610nm and 660nm, which is consistent with the simulation.

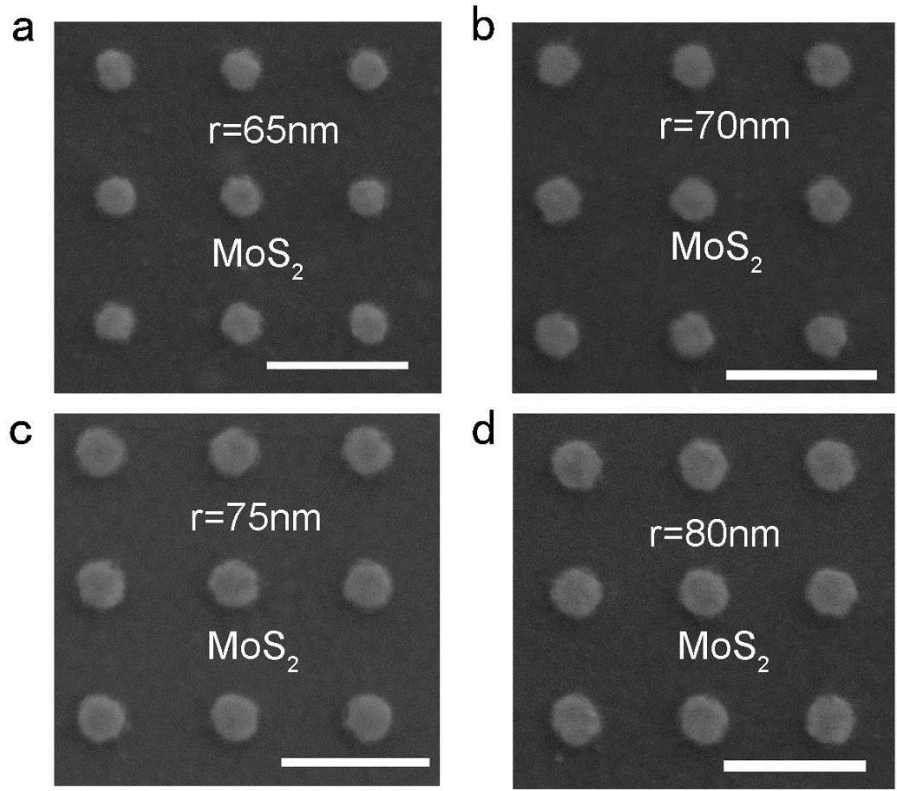

**Supplementary Figure 3.** The SEM images of MoS$_2$-Ag disk hybrid nanostructures with disk radius of (a) 65nm, (b) 70nm, (c) 75nm and (d) 80nm. The scale bar is 500nm.

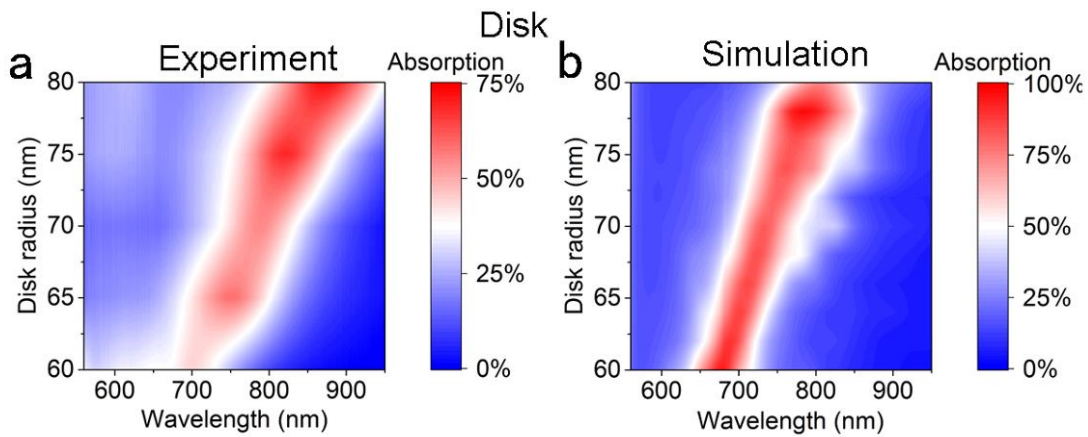

**Supplementary Figure 4.** The (a) experimental and (b) simulated (b) absorption spectra of bare Ag disk arrays on SiO$_2$/Ag substrate with disk radius ranging from 60nm to 80nm.

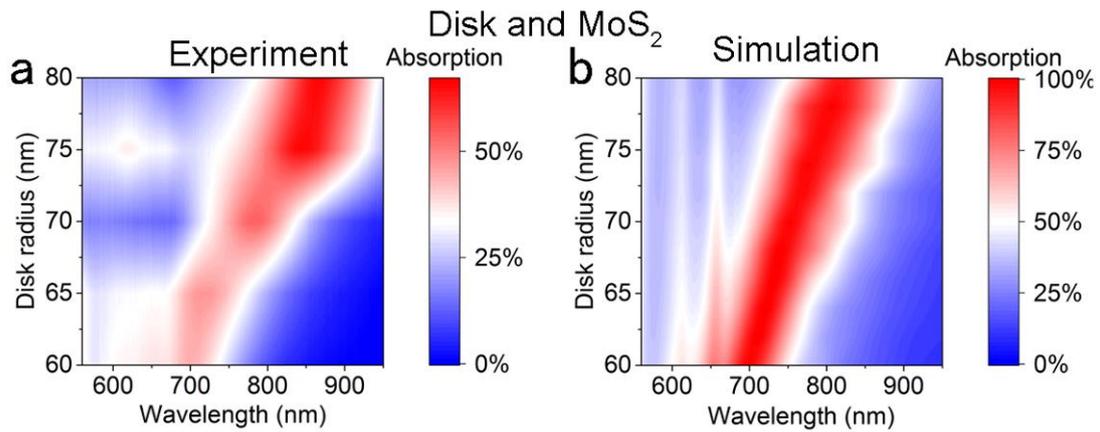

**Supplementary Figure 5.** The (a) experimental and (b) simulated absorption spectra of MoS$_2$-Ag disk hybrid nanostructures on SiO$_2$/Ag substrate with disk radius ranging from 60nm to 80nm.

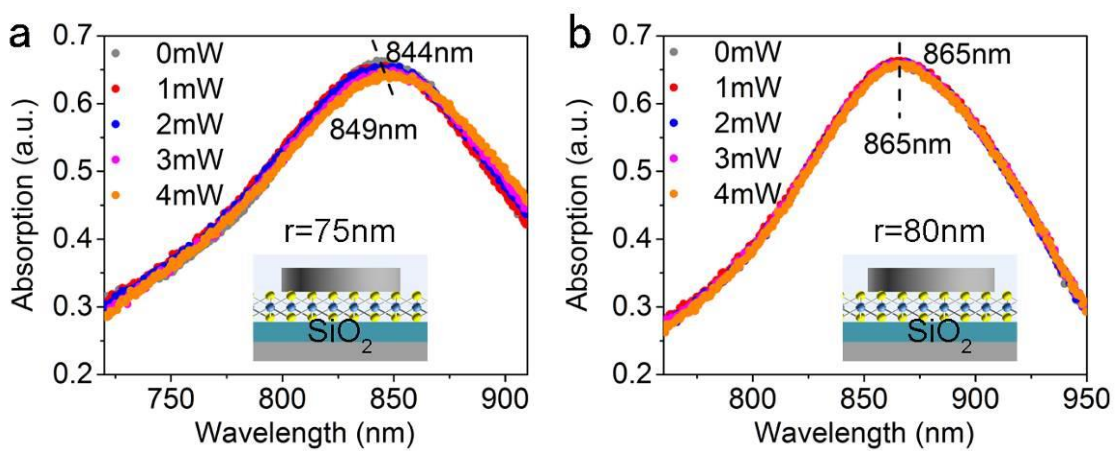

**Supplementary Figure 6.** Absorption spectra of MoS2-Ag hybrids with disk radius of (a) 75nm and (b) 80nm under various laser powers. The LSP resonance shows weak moving tendency with the laser power increasing.

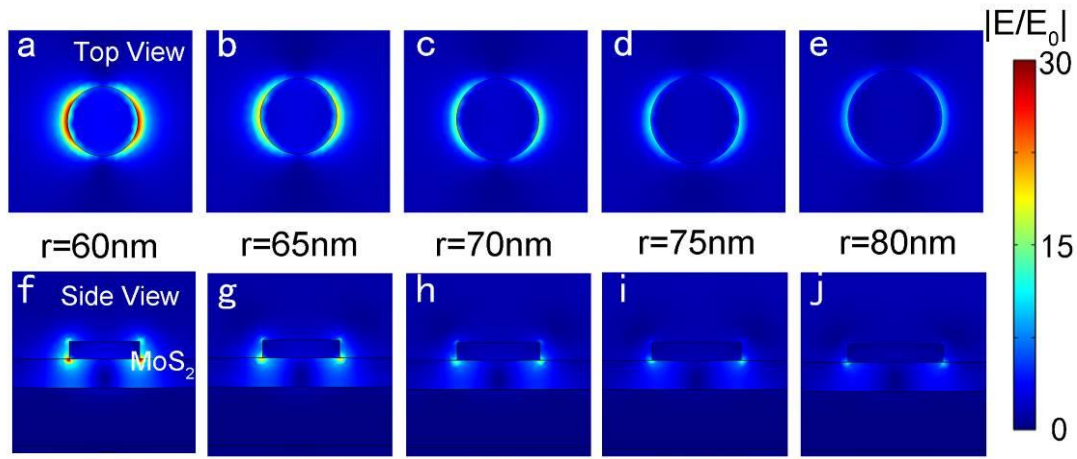

**Supplementary Figure 7.** The electric near-field distributions of different MoS2-Ag disk hybrid nanostructures on SiO2/Ag substrate with disk radius of 60nm, 65nm, 70nm, 75nm and 80nm for (a-e) top view and (f-j) side view.

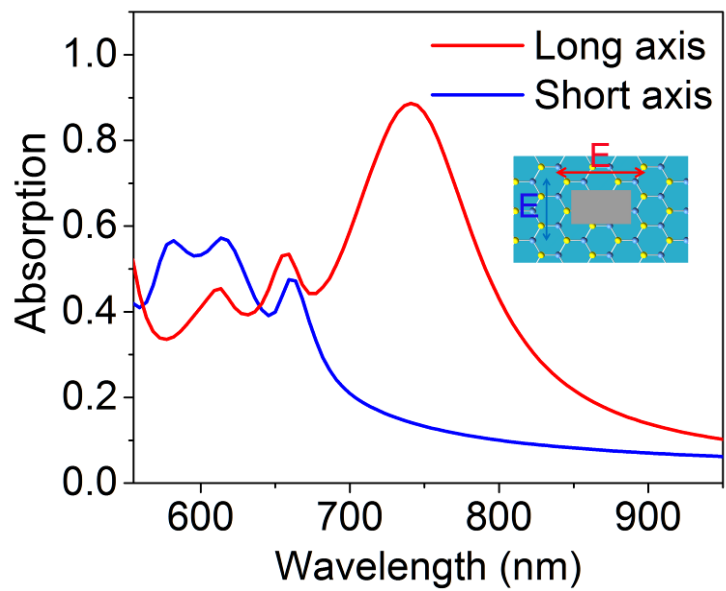

**Supplementary Figure 8.** Simulated absorption spectra of the MoS2-Ag nanorod hybrids on SiO2/Ag substrate under horizontal (red curve) and vertical (blue curve) polarization excitation.

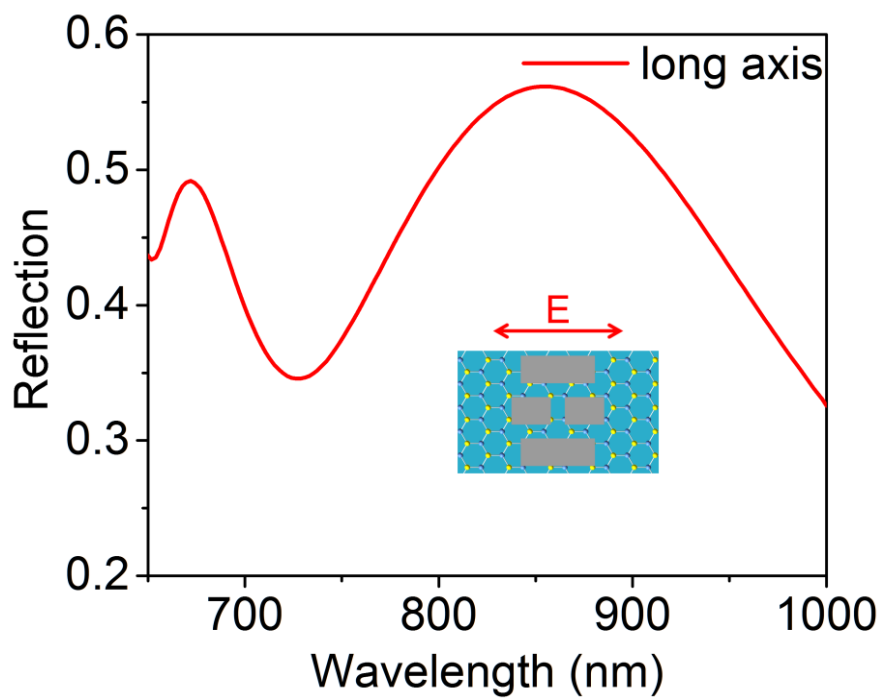

**Supplementary Figure 9**. The simulated reflection spectrum of the MoS$_2$-Ag quadrumer hybrids on SiO$_2$/Si substrate under horizontal polarization excitation.

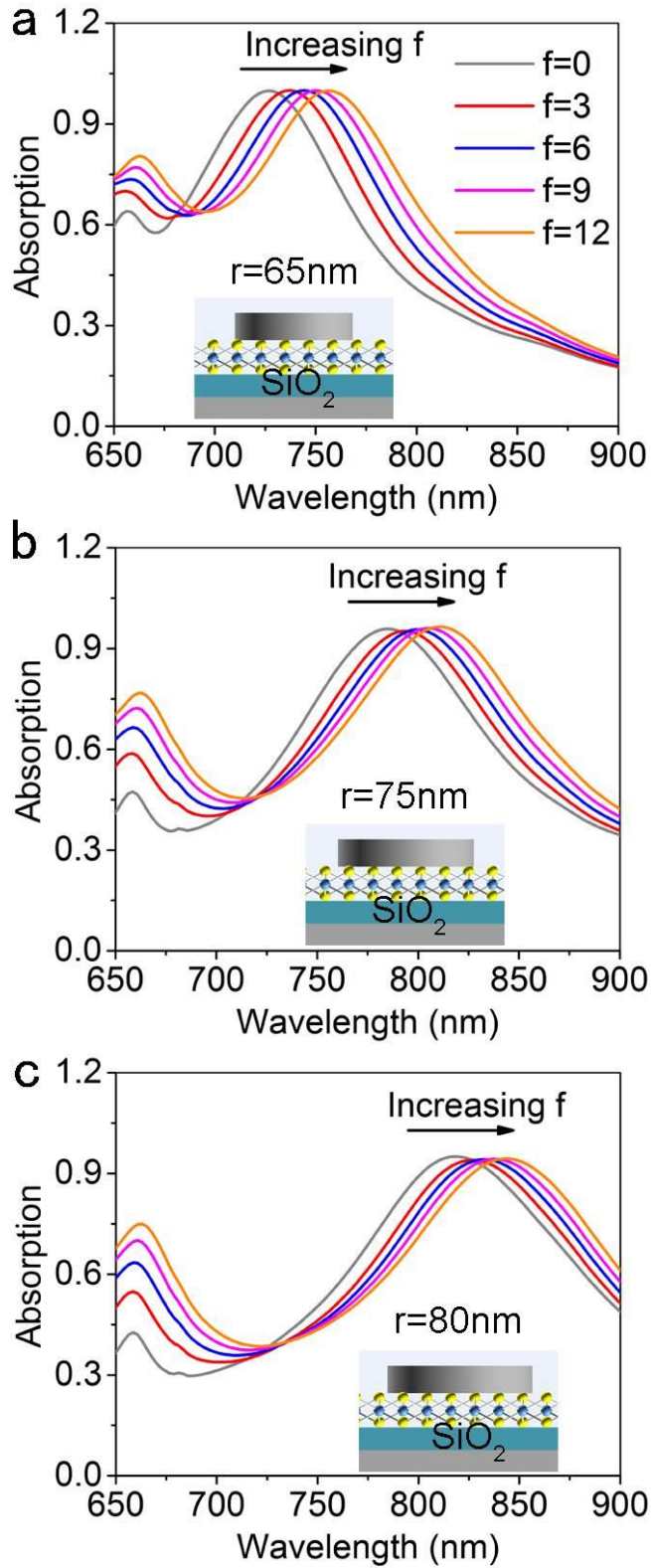

**Supplementary Figure 10.** Simulated absorption spectra of MoS2-Ag disk hybrids on SiO2/Ag substrate with disk radius of (a) 65nm, (b) 75nm and (c) 80nm by changing oscillator strength *f*.

**Supplementary Note 1: The electric near-field distributions of MoS$_2$-Ag disk hybrid nanostructures on SiO$_2$/Ag substrate with disk radius of 60nm, 65nm, 70nm, 75nm and 80nm at MoS$_2$ A- exciton absorption peak (~660 nm)**

By using the finite element method (FEM), the electric near-field distributions of MoS$_2$-Ag disk hybrids at MoS$_2$ A-exciton absorption peak (~660 nm) were simulated as plotted in the Supplementary Fig. 7 for (a-e) top view and (f-j) side view. All the disks show a dipolar plasmonic resonance with the near-field enhancement confined at the interface between the Ag disks and MoS$_2$ monolayers, which dramatically promotes the interaction between exciton and plasmon. When the LSP resonant energy is tuned close to the exciton energy (~660 nm), the near-field coupling strength of MoS2 and Ag disk is dramatically increased as shown in Supplementary Fig. 7 (a) and (f) with disk radius of 60nm and 65nm. For other Ag disks (radius of 70nm, 75 nm and 80 nm), the LSP resonance is far away from the A-exciton energy, the near-field enhancement is weak.